# SURVEY ON DATA MANAGEMENT IN RADIATION PROTECTION RESEARCH

Balázs G. Madas[1], Paul N. Schofield[2,*]

[1]Radiation Biophysics Group, Environmental Physics Department, MTA Centre for Energy Research, Konkoly-Thege Miklós út 29-33., 1121 Budapest, Hungary

[2]Department of Physiology, Development and Neuroscience, University of Cambridge, Downing Street, Cambridge CB2 3EG. UK

**The importance of datasharing is of increasing concern to funding bodies and institutions. With some prescience, the radiobiology community has established data sharing infrastructures over the last two decades, including STORE; however, the utilisation of these databases is disappointing. The aim of the present study was to identify the current state of datasharing amongst researchers in radiation protection, and to identify barriers to effective sharing. An electronic survey was prepared, including questions on post-publication data provision, institutional, funding agency, and journal policies, awareness of datasharing infrastructures, attitudinal barriers, and technical support. The survey was sent to the members of a mailing list maintained by the EC funded CONCERT project. Responses identified that the radiation protection community shared similar concerns to other groups canvassed in earlier studies; the perceived negative impact of datasharing on competitiveness, career development and reputation, along with concern about the costs of data management. More surprising was the lack of awareness of existing datasharing platforms. We find that there is a clear need for education and training in data management and for a significant programme of improving awareness of Open Data issues.**

"Scientists would rather share their toothbrush than their data!"

*Carole Goble*[1]

INTRODUCTION

There has recently been considerable discussion of the issues of data sharing and accountability in the mainstream scientific literature, amidst growing concern about the irreproducibility of reported biomedical findings, caused in part by inability to obtain primary data (1–4). Funding agencies, including the European Commission, and major journals are now implementing and enforcing increasingly stringent requirements for data sharing and transparency.

Data sharing:
- Promotes accountability for published work
- Facilitates reanalysis
- Avoids duplication
- Reduces animal use
- Stimulates new investigations
- Produces better value for money for the funding agency and the taxpayer

In response, guidelines have been agreed by major funding agencies for the sharing of data, (FAIR guidelines (5)) and between journals for transparency concerning the data underlying publications, (TOP guidelines (6)). Despite this concern, data sharing and the culture of open data are still not widespread, and there has been little investment in data sharing infrastructure and support.

**Motivation for this study**

Since 2002 many surveys have been carried out on attitudes to, and experiences of data sharing, which paint a pessimistic picture of the administrative and cultural challenges. In radiation protection research, there are well-established infrastructures for data sharing, but these databases are still underused by researchers. The objectives of the present study were to find out why researchers are not sharing their data and what kind of support would be effective to encourage population and utilisation of existing sharing platforms.

---

[1] https://doi.org/10.1093/bib/bbn003

*Corresponding author: balazs.madas@energia.mta.hu

METHODOLOGY

The CONCERT project (http://www.concert-h2020.eu/en), supported by the European Commission, functions as an umbrella organisation for the integration of radiation protection research across Europe, and therefore provides a large constituency of scientists and regulators in many areas of radiation safety and research. An electronic survey was prepared which included questions on accessibility of respondents' most recent article and its data, institutional, funding agency, and journal policies, awareness of data sharing infrastructures, major barriers, and desired support for data sharing. The survey was sent to the members of the broadest mailing list maintained by CONCERT requesting them to distribute it extensively. 46 complete responses were received, mainly from respondents describing themselves as senior investigators.

RESULTS

Table 1 provides a general overview of the respondents' attitude towards data sharing. All the raw data are available within, or associated with, the most recently published article of 24% of the respondents. About 55% are open to share raw data with people who will use the data in a responsible way. More than 20% share data only with co-authors.

**Table 1. Accessibility of the raw data underlying the most recently published article of the respondent**

| Answer choices | Responses |
| --- | --- |
| All the raw data is available in the article. | 9.09% (4) |
| All the raw data is available either in the article or in the supplementary material. | 15.91% (7) |
| The raw data not published along with the article are available in a public database like STORE. | 0.00% (0) |
| There are raw data that are accessible only for the authors, but I would make the data available to anyone who convincingly argues they will use them in a responsible way. | 54.55% (24) |
| There are raw data that are accessible only for the authors. | 18.18% (8) |
| There are raw data that are inaccessible even to co-authors. | 2.27% (1) |

Figure 1 shows that many institutions and funding agencies (about 70%) do not have a mandatory data sharing policy. In addition, even when investigators are aware of mandatory data sharing policies, only half fully comply (not shown).

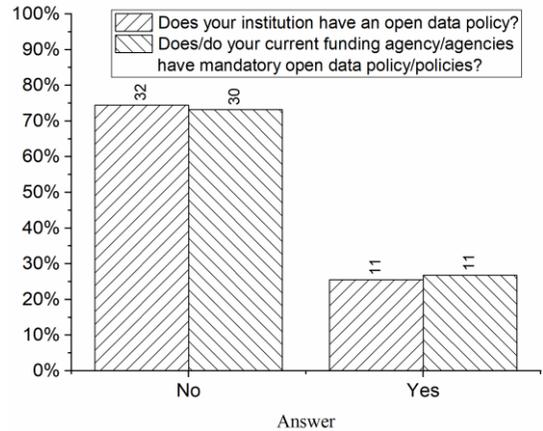

Figure 1. Open data policies applying on the respondent.

The limited awareness of the respondents can also be seen in Figure 2 showing that about two thirds are not aware of any data repository that accepts all of the data types they generate.

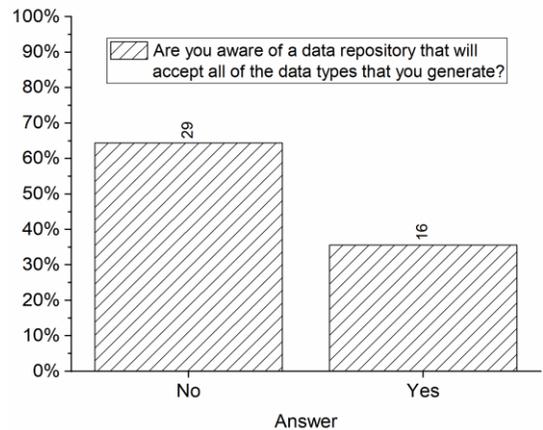

Figure 2. Awareness of data repositories accepting data generated by the respondent.

Figure 3 shows the most surprising result, that 44% of the respondents had not heard about the STORE database (www.storedb.org), which is one of the most important data sharing infrastructures in European radiation biology. While FREDERICA (7), much older and dedicated to the environmental community, was slightly better known (40% had heard about it, and 21% downloaded data from it), awareness of ERA (8), JANUS (9), JRA, and NRA is much lower. All of these databases have been extensively reported at meetings and most in the mainstream literature.

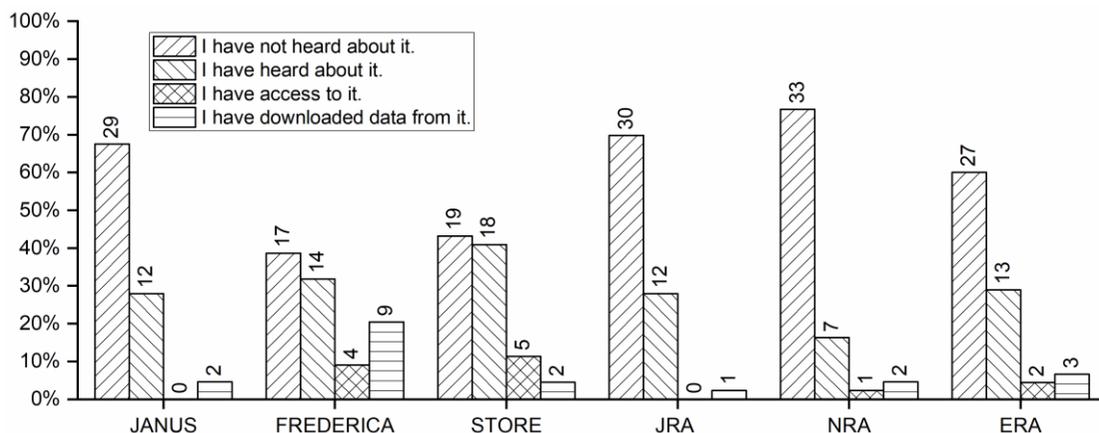

Figure 3. Awareness of extensively reported databases in radiation biology: JANUS - Janus Tissue Archive, FREDERICA - Radiation Effects Database, STORE - Sustaining access to Tissues and data from Radiobiological Experiments, JRA - Japanese Radiobiology Archives, NRA - National Radiobiology Archives, ERA - European Radiobiology Archives.

Table 2 summarizes the reasons given for refusing to share data or for conditional sharing of published data. Here (as well as in Table 3) the respondents marked how much they agreed with the statements on a five level scale: (1) strongly disagree, (2) rather disagree, (3) neither disagree, nor agree, rather agree (4), strongly agree (5). The table shows the weighted averages.

**Table 2. Reasons given for refusing to share data or for conditional sharing of published data**

| What are the major barriers which prevent you making your raw data publicly available? | Weighted average |
|---|---|
| I do not have any barriers. | 3.09 |
| I am afraid that my data will be used without giving credit to me. | 3.12 |
| I am afraid that I lose control of my data. | 3.15 |
| My institution does not allow me to share all data. | 2.67 |
| My co-authors do not allow me to share all data. | 2.66 |
| I do not have time to make my raw data understandable to everyone. | 3.56 |
| I do not see any advantage of sharing data. | 1.95 |
| I do not have funding to make data publicly available. | 3.89 |
| I am not allowed to share data due to human data protection and consenting constraints. | 3.39 |
| There is a lack of suitable repository. | 2.66 |
| I do not know how to share data. | 3.08 |
| I have concerns about scientific competition. | 3.03 |

Table 3 shows what kind of support would increase the chance that the respondent would be made the raw data publicly available. It can be seen that many researchers are not aware of databases and the intellectual property rights related to their raw data, so they need more information on these. Besides this, respondents would like to get internal or external help in preparing the data for public sharing.

**Table 3. Kind of support potentially helping data sharing**

| What kind of support would increase the chance that your raw data would be made publicly available? | Weighted average |
|---|---|
| Specific funding for covering working hours required for data sharing. | 3.91 |
| More information about databases and my rights related to my raw data. | 4.09 |
| If my co-authors would make our raw data publicly available. | 3.48 |
| Internal assistance (provided by my institution) who helps in making data publicly available. | 3.96 |
| External assistance (provided by the database administrator) who helps in making data publicly available. | 3.77 |
| Specific requirements from funding bodies making public data sharing obligatory. | 3.61 |

CONCLUSIONS

The analysis of our results demonstrates that members of the radiation protection research community behave rather similarly to other communities surveyed in the face of a change of culture regarding data sharing.
The reasons given for failure of investigators to share data are congruent with previous studies in other domains of the biological sciences; fear of competition, desire to have unique access for career development reasons, fear of others "misinterpreting the data" and an underlying insecurity about their own analysis. Added onto this are issues of the costs and training required for routine data sharing, lack of understanding of licensing conditions, the advantages of data sharing for the originator, and the value of data reuse. It is surprising that given the level of discussion across the community there seems to have been only incremental movement in attitudes towards best practice in data management over the past decade (10–13) although one study (14) did show the beginnings of movement amongst younger scientists. This attitudinal problem is not helped by papers in major journals in supporting the right of authors to keep their primary data secret (15–17) and the failure of many journals either to require open data access or to enforce their own policies.

What was surprising was lack of knowledge of existing and well established – and cost-free - sharing infrastructures in the radiation sciences and a lack of policies amongst funding agencies, institutes and universities across Europe. Unfamiliarity with many of the issues was also something of a surprise, given the high profile discussions in the literature over recent years. That this is to a degree a generational issue also seems to emerge, with younger scientists more ready to consider making their data open, suggesting that even when career development is a critical issue, a new cultural norm is developing.

We find that there is a clear need for education and training in data management and a significant programme of increasing awareness of Open Data issues is needed, consistent with the developing policies in Horizon 2020 and those already adopted by major funding agencies such as the NIH. Declarations of clear and mandatory policies for data sharing need to be matched by guidance, monitoring and most importantly funding for data management, training, and sustainability of sharing platforms themselves. The cost of doing so is unlikely to be high; the cost of not doing so is likely to be much higher, both to the public purse and, because of the issues of reproducibility, to the reputation of radiation biology in society and the authority of the radiation protection establishment.

## ACKNOWLEDGEMENT

The research leading to these results was funded from the EURATOM research and training programme 2014–2018 under Grant agreement no. 662287 (CONCERT).